\documentclass[twocolumn,apj,iop]{openjournal}

\usepackage{listings}

\DeclareRobustCommand{\VAN}[3]{#2}
\let\VANthebibliography\thebibliography
\def\thebibliography{\DeclareRobustCommand{\VAN}[3]{##3}\VANthebibliography}

\usepackage{graphicx}	
\usepackage{amsmath}	

\usepackage{orcidlink}

\usepackage{caption}

\shortauthors{R. Jackim et al.}

\begin{document}
\lstset{language=SQL}

\title[GALEX-Gaia White Dwarfs]{The GALEX-Gaia-EDR3 Catalogue of Single and Binary White Dwarfs}
\shorttitle{GALEX-Gaia White Dwarfs}
\author{\vspace{-1.5cm}
Ryan Jackim\orcidlink{0000-0002-5064-2760}$^1$}
\author{Jeremy Heyl\orcidlink{0000-0001-9739-367X}$^1$}
\author{Harvey Richer\orcidlink{0000-0001-9002-8178}$^1$}

\affiliation{$^{1}$Department of Physics and Astronomy, University of British Columbia, Vancouver, B.C. V6T 1Z1, Canada}

\begin{abstract}
We present a catalogue of white dwarf candidates constructed from the GALEX and Gaia EDR3 catalogues. 
 The catalogue contains 332,111 candidate binary white dwarf systems and 111,996 candidate single white dwarfs.  Where available, the catalogue is augmented with photometry from Pan-STARRS DR1, SDSS DR12 and classifications from StarHorse. We fit photometric data with modeled white dwarf cooling sequences to derive mass, age and effective temperature of the white dwarf as well as mass estimates for the companion.  We test our classifications against StarHorse, the Gentile-Fusillo Gaia EDR3 catalogue, and white-dwarf-main-sequence binaries identified in SDSS DR12.  This catalogue provides a unique probe of the binarity of white dwarfs as well as the abundance of white-dwarf giant binaries and large mass-ratio stellar binaries which are difficult to probe otherwise.
\end{abstract}

\keywords{
White Dwarf -- Binaries: general -- Catalogues
}
\maketitle


\section{Introduction}

White dwarfs (WD) are stellar remnants originating from stars with mass less than 8 $\textrm{M}_\odot$ which have reached the end of their evolutionary track. 
White-dwarf-main-sequence (WDMS) binaries are stellar systems with a WD that has formed from the initial primary star and a main sequence (MS) star. Similarly, white-dwarf-giant (WDG) binaries are systems where a WD has formed from the initial primary star and the secondary is now a giant star.  
We will refer to both types of systems generically as white-dwarf-non-degenerate (WDND) binaries (in contrast with double-degenerate, DD, binaries).  WDs have low luminosity in the optical. When paired with a MS star more massive than a G star, it is difficult to identify the WD in the system due to the comparatively high luminosity of the companion. WDND systems offer the ability to construct studies on comprehensive binary population synthesis as well as insights into pre-cataclysmic variables \citep{2004A&A...419.1057W}. 
The evolution of WDND systems is largely controlled by the separation between the two stars. 
Three-quarters of WDND systems have orbital periods wide enough that the progenitor of the WD evolved independently of the other star in the system. Typically these systems will have an orbital period longer than 400 days \citep{2004A&A...419.1057W}.   
The remaining twenty five percent are close enough to each other to exchange mass during their stellar evolution. In these close binary systems, the more massive companion evolves off the main sequence first and moves onto the red giant branch (RGB). The RGB's outer atmosphere  
envelops the lower mass companion star and transfers mass. This is referred to as the common envelope (CE) stage. The RGB star's envelope is eventually ejected and it evolves into a WD creating a WDND system \citep{2013A&ARv..21...59I}. 

The atmospheres of WDs are generally found as either hydrogen-rich (DA) or helium-rich (DB). The fraction of WDs with helium-rich atmospheres increases significantly when $T_\text{\scriptsize eff}$ of the WD is below 10,000 K. 
The reason for this change is likely caused by convection mixing the upper hydrogen atmosphere with the more massive helium envelope beneath. It is expected that this convection process will turn all hydrogen-rich atmosphere WDs into helium-rich atmosphere WDs \citep{1997ApJS..108..339B}. 

There have been several previous surveys that have attempted to gather samples of the WDND systems. The Sloan Digital Sky Survey (SDSS) identified WDND binary systems from Radial Velocity (RV) measurements of optical spectra. Using this technique, just over 3,300 WDND binaries have been identified \citep{2016MNRAS.458.3808R}. SDSS was primarily developed to identify quasars and galaxies, leading to most of the WDND systems identified being the same colour as quasars. This creates a bias towards hot ($\gtrsim$10,000-15,000 K) WDs with M-dwarf spectral type (>M3–4) companions \citep{2016MNRAS.458.3808R,2013MNRAS.433.3398R}. A correction done by \cite{2013MNRAS.433.3398R} attempted to use photometric data to identify WDND systems with colder WDs. Of the 3,419 candidates selected, 71\% of the sources were expected to have a cool ($\lesssim$ 10 000-15 000 K) WD and a M-dwarf secondary ($\sim$M2-3). 
A similar method was employed by the Large Sky Area Multi-Object Fiber Spectroscopic Telescope (LAMOST) Survey which has identified 876 WDND systems \citep{2018MNRAS.477.4641R}. 
These previous studies of WDND systems have only observed WDND systems with M-dwarf companions. A study done by \cite{2021MNRAS.506.5201R} reported 112 WDND binary systems within 100 pc, located in the gap between the WD and MS region in the visual CMD. Their study did not include WDND systems lying within the MS region where the MS star dominates the light of the WD. They were able to conclude that their identification only accounted for roughly 9\% of WDND systems and that 91\% of WDND binaries remain obscured in the MS region. Thus, it is imperative that we identify alternative methods for assessing WDND systems enhance our understanding of the evolution of binary systems.

To attempt to identify WDND systems, we used several photometric survey's that contain optical and UV data. 
Two primary catalogues were utilized. The Galaxy Evolutionary Explorer (GALEX) is a NASA explorer mission designed to carry out a space-based UV sky survey. The GALEX survey allows for UV detections of objects as faint as 23rd magnitude \citep{2005ApJ...619L...1M}. WDs with a temperature range of 30,000 K to less than 4,500 K will emit in the UV 
\citep{2019ApJ...876...67B}.  This makes the GALEX survey an excellent tool in which to identify WD systems. Gaia Early Data Release 3 (EDR3) \citep{2021A&A...649A...1G} is a catalogue containing 1.8 billion stars limited to objects brighter than 21st G-magnitude. It contains improved photometry relative to Gaia Data Release 2 and detailed distance measurements of objects through observation of their parallactic shift. Gaia EDR3 contains the same astrometric data as Gaia DR3; therefore, the third data release does not add additional data for the match. The combined use of both these catalogues allows us to detect WDs through GALEX and collect both photometric and distance information of the systems through Gaia EDR3. 

Using WD stars identified in GALEX and Gaia EDR3, we constructed a catalogue containing a total of 444,107 candidate single and binary WDs. Of these, 111,996 candidates match to WD objects identified by \cite{2021MNRAS.508.3877G}. The remaining 309,584 are candidate WDND systems where the GALEX UV photometry reports values similar to WD, but Gaia EDR3 reports values similar to MS or giant stars. This was accomplished by identifying all GALEX entries near the WD sequence in the UV and cross-matching them with Gaia EDR3 entries. We then attempted to gather stellar parameter estimates for both the WD and non-degenerate secondary by applying a two-star fitting algorithm based on the stellar isochrones of \cite{2012MNRAS.427..127B} and the WD cooling models of \cite{2020ApJ...901...93B}. We analyzed the resulting masses for all the stars in the system and reported our estimated masses in our catalogue. Mass estimates from the StarHorse MS catalogue \citep{2021arXiv211101860A} and \cite{2021MNRAS.508.3877G} WD catalogue were included to supplement our estimates.

\section{Catalogue Construction}

\subsection{Data Selection}
\begin{figure*}
    \includegraphics[width=0.32\linewidth]{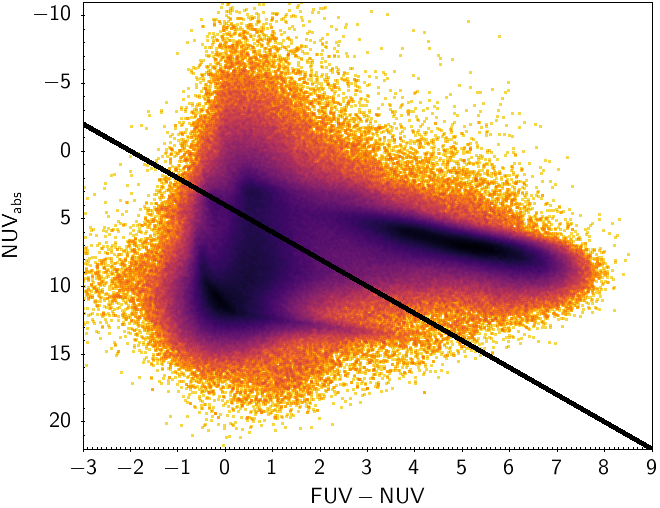} 
    \includegraphics[width=0.32\linewidth]{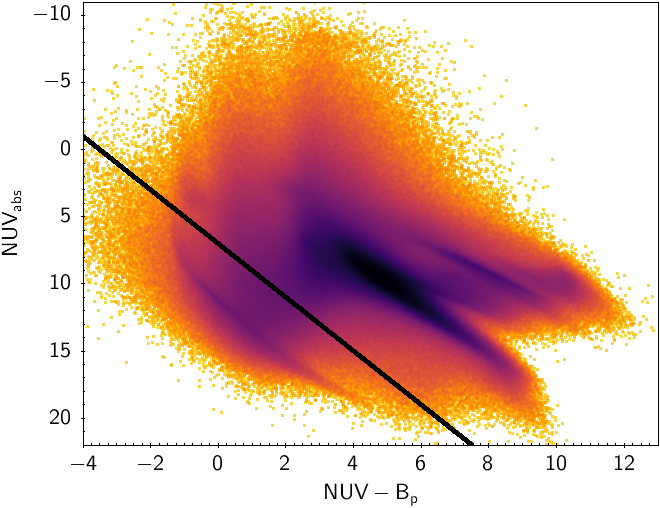}
    \includegraphics[width=0.32\linewidth]{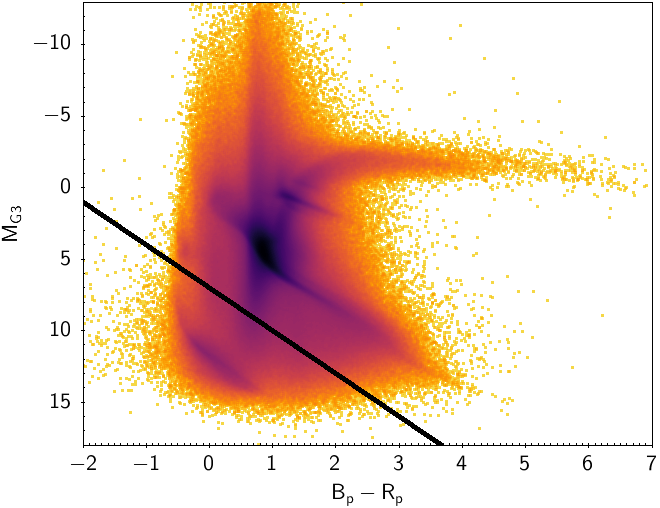}
    \caption{The distribution of GALEX-Gaia EDR3 cross-match is shown in the above three figures. The left image depicts the objects that succeeded our cross-match in the GALEX FUV-NUV filters. The middle figure depicts the same objects in the GALEX NUV-Gaia Bp filters. The right figure depicts these objects in the Gaia Bp-Rp filters. The black line represents our cut described in ADQL~\ref{lst:photo}. Objects that lie below the black line in any of these panels are considered WDs or WDNDs and included in our catalogue. 
    }
    \label{fig:image2}
\end{figure*}

The procedure to select WDND systems is similar to the \cite{2021MNRAS.508.3877G} selection of WD systems in the Gaia EDR3 data set. We sought to expand our list of potential WDND identifications by including likely UV-WDs as one of the components in the binary system. 
 To begin, we produced a cross-match between GALEX and Gaia EDR3 using Vizier. The purpose behind this cross-match was to identify stellar systems with UV excesses that resembled WD stars. If these systems had an excess similar to a WD, it is likely that there is a WD present. We then implemented a broad cut which defines an area of the colour-magnitude diagram (CMD) space where WDs or WDND systems are located. At this stage, the cross-match was focused on gathering a wide selection of potential objects with little consideration for contaminants:
 \begin{lstlisting}[caption=Photometric Cut,label=lst:photo]
NUVmag-DM-0.6*(FUVmag-NUVmag) > 8 OR
NUVmag-DM-
  2*(NUVmag-phot_bp_mean_mag) > 8 OR
phot_g_mean_mag-DM-3*bp_rp > 8
\end{lstlisting}
where $\texttt{DM}$ is the distance modulus determined using the Gaia EDR3 parallax.
Our selection focused on locating white-dwarf-like objects within one of three wavelength regions. These three regions are: the far-UV (FUV) and near-UV (NUV) defined exclusively by the GALEX data-set in the left plot of Figure \ref{fig:image2}, the near-UV from GALEX to the Gaia Bp filter described in the center plot of Figure \ref{fig:image2}, and the Gaia EDR3 visual magnitudes described by the Bp and Rp filters shown in the right plot of Figure \ref{fig:image2}. The black line in Figure \ref{fig:image2} depicts a cut where everything below the line is considered a WD or a WDND. Everything above the line is excluded from our catalogue. If an object was below the black line in one of the wavelength regions, it was considered a WD or WDND regardless of whether it succeeded in the other two regions. 

A large fraction of our collected data had unreliable astrometric or photometric data. We apply the quality cuts in ADQL \ref{lst:correct} to our sample, 
which excluded non-WD contaminants and entries with poor photometry. 
\begin{lstlisting}[caption=Error Cut,label=lst:correct]
parallax_over_error >= 4 AND
    (astrometric_excess_noise_sig < 2 OR
    astrometric_excess_noise < 1.5 OR
    astrometric_params_solved < 32)
\end{lstlisting}

 Because of the potential presence of a companion star contaminating the photometry of the WD, the quality cuts were generous enough to allow for this circumstance. 

Two additional cross-matches were done to identify if the objects had data from SDSS DR12 or Pan-STARRS DR1. If the objects matched SDSS or Pan-STARRS entries, the data were included in the catalogue. Those that did not were left blank. 

\begin{figure}
	\includegraphics[width=\columnwidth]{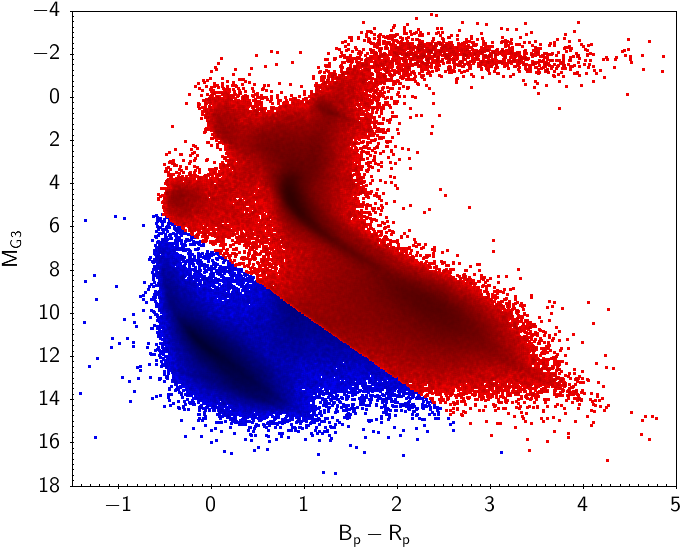}
    \caption{Gaia absolute G magnitude and Bp-Rp of the 444,107 entries in the catalogue. The blue region represents objects that lie within or near the WD region in Gaia and the red region represents objects identified as a WD in the GALEX filters. The blue region contains 134,523 entries while the red region contains 309,584 entries.}
    \label{fig:cmd}
\end{figure}

\begin{figure}
	\includegraphics[width=\columnwidth]{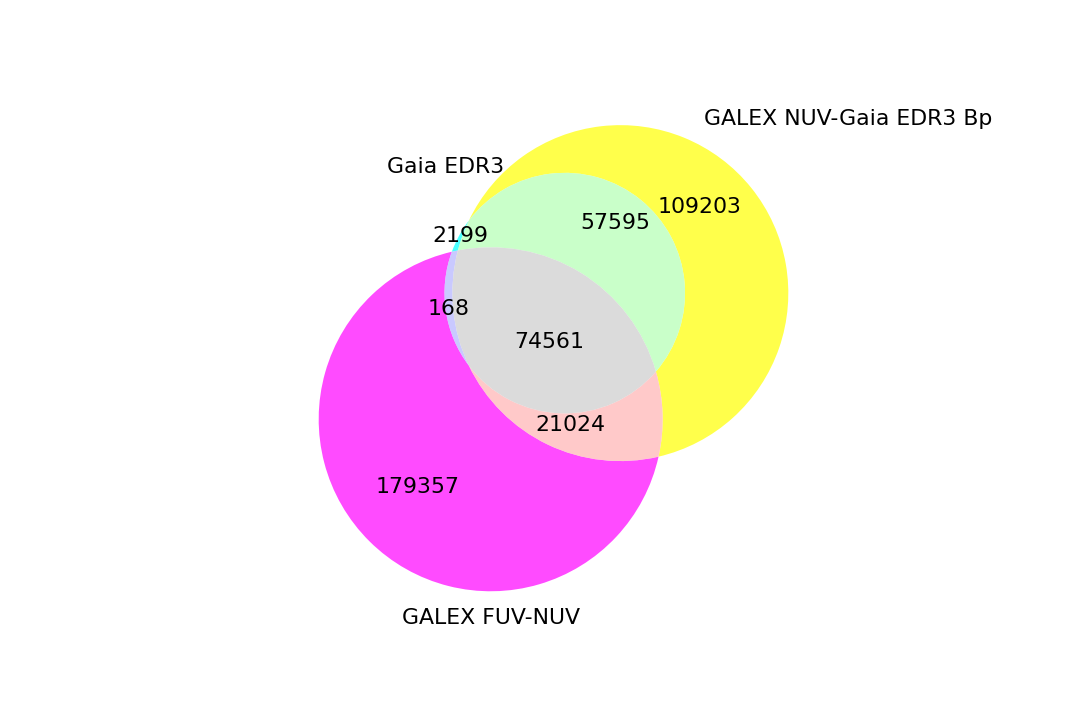}
    \caption{This figure describes the number of entries in our catalogue that survived at least one of our cuts in the colour-magnitude diagram. The cyan slice (2,199 stars) indicates the number of objects that appear as WDs only in the Gaia CMD and not the GALEX NUV-Gaia Bp or GALEX FUV-NUV CMD. The yellow circle describes the number of objects only identified as WDs in GALEX NUV-Gaia Bp filters and not the Gaia Bp-Rp or GALEX FUV-NUV CMD. The magenta circle illustrates the number of objects only identified as WDs in the GALEX FUV-NUV filters and not the Gaia Bp-Rp or GALEX NUV-Gaia Bp CMD. The intersections of the circles indicate objects which have two or more WD identifications. The 168 objects identified as a WD only in Gaia EDR3 and GALEX FUV-NUV CMD and not in the GALEX NUV-Gaia Bp CMD, are depicted in darker purple (168). Over half of the objects that we identify as WDs are not identified as WDs in the Gaia ED3 filters, indicating that they appear as non-WD-like objects in the Gaia EDR3 filters.}
    \label{fig:venn_all}
\end{figure}

After quality cuts were applied, 444,107 possible WD systems remain. Figure \ref{fig:cmd} describes the distribution of these 444,107 objects in the Gaia EDR3 photometric filters. The blue region signifies objects that were similar to WDs in Gaia EDR3 and the red signifies those that did not look like WDs in Gaia EDR3, but were similar to WDs in other wavelengths. Upon inspection of Figure \ref{fig:cmd}, it is clear that there is a large variety of objects identified as WD across the MS and the evolved branch. The total numerical distribution of objects that succeeded our three different WD cuts can be seen in the weighted Venn diagram of Figure \ref{fig:venn_all}. The magenta circle depicts objects that look like WDs only in the GALEX FUV-NUV filters and not WDs in the GALEX NUV-Gaia Bp or Gaia Bp-Rp filters. The yellow circle describes objects that look like WDs only in GALEX NUV-Gaia Bp and not WDs in the GALEX FUV-NUV or Gaia Bp-Rp filters. Finally, the cyan slice (2199 stars), defines the number of objects that were identified as a WD in only the Gaia filters and not WDs in the NUV-Bp or FUV-NUV filters. A majority of the objects in the cyan slice, did not contain FUV data or were objects that existed above the WD sequence in the Gaia filters. If one of the circles has been overlapped by another, then the object looks like a WD in two filters. The central section where all three circles overlap describes the number of objects that look like WDs in all three filters. The cyan Gaia EDR3 circle in Figure \ref{fig:venn_all}, including all overlapping sections, directly depicted the blue region in Figure \ref{fig:cmd}, totaled 134,523 objects. The yellow, pink and magenta circles correspond to objects that were identified as WDs in GALEX UV but not Gaia optical bands and totaled 309,584 objects. The red region in Figure \ref{fig:cmd} corresponds to these 309,584 objects. By observing this red region in Figure \ref{fig:cmd}, it is clear that there are many objects that appear to be MS or giant stars that look like WDs in the GALEX UV. Thus, we consider these 309,584 objects to be candidate WDND systems. 

A number of entries in Figure \ref{fig:cmd} appear to lie within the WD sequence in the Gaia filters. To attempt to identify these possible single WDs, a cross-match was done with the \cite{2021MNRAS.508.3877G} WD catalogue. Of the 444,107 entries, 111,996 matched to WDs previously identified by \cite{2021MNRAS.508.3877G}. A total of 332,111 objects remain unmatched.

\subsection{Extinction}

We estimate a value for the reddening in Gaia EDR3 and GALEX filters using a method similar to \cite{2019MNRAS.482.4570G}. An estimated Galactic extinction is identified from the 
\cite{1998ApJ...500..525S} dust map for each object individually. The extinction coefficients for the GALEX filters were derived by \cite{2019MNRAS.489.5046W}, while coefficients similar to \cite{2020ApJ...901...93B} were utilized for the Gaia EDR3 filters. Formulae for absolute magnitude given the apparent magnitude, distance modulus and extinction (assumed uniform along the line of sight) are shown below. 

\[
G_\textrm{\!\scriptsize star} = G_\textrm{\!\scriptsize obs} - DM - 0.835 A_V\left [ 1-\exp \left ({-\frac{\sin{|b|}}{200\varpi}} \right ) \right ]
\]
\[
G_\textrm{\scriptsize Bp,star} = G_\textrm{\scriptsize Bp,obs} - DM - 1.364A_V\left [ 1-\exp \left ({-\frac{\sin{|b|}}{200\varpi}} \right ) \right ]
\]
\[
G_\textrm{\scriptsize Rp,star} = G_\textrm{\scriptsize Rp,obs} - DM - 0.778 A_V\left [ 1-\exp \left ({-\frac{\sin{|b|}}{200\varpi}} \right ) \right ]
\]
Here, $G_\textrm{\!\scriptsize obs}$ is the observed magnitude, $A_V$ is the individual Galactic extinction identified from the \cite{1998ApJ...500..525S} dust maps in the $V-$band, $b$ is the Galactic latitude, and $\varpi$ is the parallax in arcsec. Further details concerning this function can be found in \cite{2019MNRAS.482.4570G}. 

\subsection{Characterization of WD Single and WDND Systems}

Using the processed photometry of the systems, an estimate of the WD mass and age was attempted. WD candidates identified below the Gaia EDR3 optical cut are assumed to be WDs without a companion. These single WD stars were characterized using WD cooling models developed by \cite{2020ApJ...901...93B}. 
Using \cite{2020ApJ...901...93B} WD cooling models and initial estimates for the mass and cooling age, photometric models of the system in all potential filters were constructed. 
These photometric models were compared to the photometric data in a $\chi^2$ test. The mass and cooling age of the WD was allowed to change in order to minimize the $\chi^2$. 
The processed photometry and modeled photometry used GALEX FUV NUV bands, Gaia EDR3 bands and all the SDSS filters available. Where data were not present, or a value of 0 was reported, the band or filter data were ignored in the $\chi^2$ minimization. The mass of the secondary star for these objects is reported as 0 in the case of single WDs. The estimated mass of the WD was limited by the models to be within 0.2-1.3 $\textrm{M}_\odot$. The estimated masses and cooling ages were applied to the \cite{2020ApJ...901...93B} WD cooling models to develop an estimate for the temperature of the WD.

For the WD binary candidates, a similar method $\chi^2$ minimization fit was carried out. However, we modified the minimization to attempt to fit a WD mass, a WD cooling age and a MS or giant secondary mass by adding the modeled photometry of the WD with the modeled photometry of a MS or giant star. Padova isochrones \citep{2012MNRAS.427..127B}
were used to model the MS or giant photometry for the fit. Due to the presence of both high-mass stars and evolved stars in the sequence, two different isochrones were chosen to model the data. A cut was done in the optical regions to divide the evolved stars from the rest of the MS objects. The evolved stars were fitted to an isochrone with an age of 10 billion years. It is assumed that all the evolved stars fit on this one isochrone in order to get a rough estimate of mass for these stars. All the stars in this catalogue were cross-matched with the StarHorse catalogue \citep{2021arXiv211101860A} to provide an additional estimate for the mass of the secondary star.

A mass for the single WD fit and binary WD of 0.3 $\textrm{M}_\odot$ was used for an initial mass estimate. The expected mass of most WDs is between 0.4-0.6 $\textrm{M}_\odot$. However, a smaller initial mass was used to attempt to identify incidences where the fitting function failed to deviate from the initial mass provided. Figure \ref{fig:wd_dist} reveals no such failure to deviate, whereas WDND WD mass estimates showed a significant number of entries that failed to deviate from the initial mass. 

A correction had to be applied to our data while attempting to fit GALEX NUV filter for candidate binary systems. When the best fit MS or giant star was chosen out of Gaia EDR3 filters, the modeled GALEX NUV magnitude for the MS or giant star was smaller than the NUV filter data. This means that the secondary star is emitting more NUV light than we expect for a MS or giant star of this type. We attribute this anomaly to poor MS and giant photometry models in the GALEX NUV and exclude the NUV filter in our fits where the anomaly is present.

This catalogue reports the information present in the GALEX and Gaia EDR3 cross-match as well as SDSS DR12 and Pan-STARRS DR1 data, if any are present. It also reports the cross-matched StarHorse catalogue entries as well as the estimated WD mass, WD cooling age and MS star's mass for both DA and DB atmosphere WDs.

\section{Catalogue Contents}  

The catalogue contains 444,107 unique candidate WD systems. These systems were identified as candidate WDs using GALEX and Gaia EDR3 data. The first section contains identification information for each star system (Right Ascension and Declination). GALEX data follows the identification section. The GALEX portion of the table contains all the GALEX data for each entry. The Gaia EDR3 section contains information regarding each star based on the Gaia EDR3 data. Pan-STARRS and SDSS DR12 data are listed next. Entries will be empty if the star did not have either Pan-STARRS or SDSS DR12 data to correspond with the Gaia EDR3 data. Entries that were cross-matched to StarHorse \citep{2021arXiv211101860A} follow after the SDSS and Pan-STARRS data. The entries having no StarHorse data were left blank. Entries that matched with the \cite{2021MNRAS.508.3877G} WD catalogue are listed after StarHorse. Following this section are three boolean columns that are true if the star passed our photometry cut for either the GALEX FUV-NUV (fuvwd), GALEX NUV - Gaia EDR3 Bp (nuvwd) and Gaia EDR3 (gaiawd). The final section of the catalogue is the mass estimate of the WD and the non-degenerate star for both DA model estimates and DB model WD estimates. The secondary star column will have an entry of 0 for WDs that are assumed to be without a companion. Entries without a value were reported as 99.9 or 99.0. See Appendix A for a detailed list of all the columns. 

\section{Analysis}

\subsection{Single White Dwarfs}

\begin{figure}

	\includegraphics[width=\columnwidth]{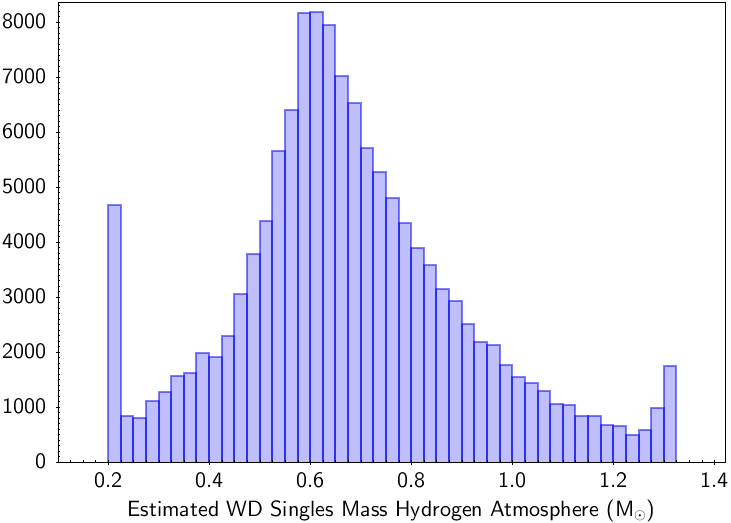}
    \caption{The estimated mass distribution of WDs identified as WD stars in Gaia filters assuming hydrogen atmospheres (DA models). The sudden spikes on the tails of the distribution are likely WDNDs mixed into the single WD data and did not properly fit our models.}
    \label{fig:wd_dist}
\end{figure}

\begin{figure}

	\includegraphics[width=\columnwidth]{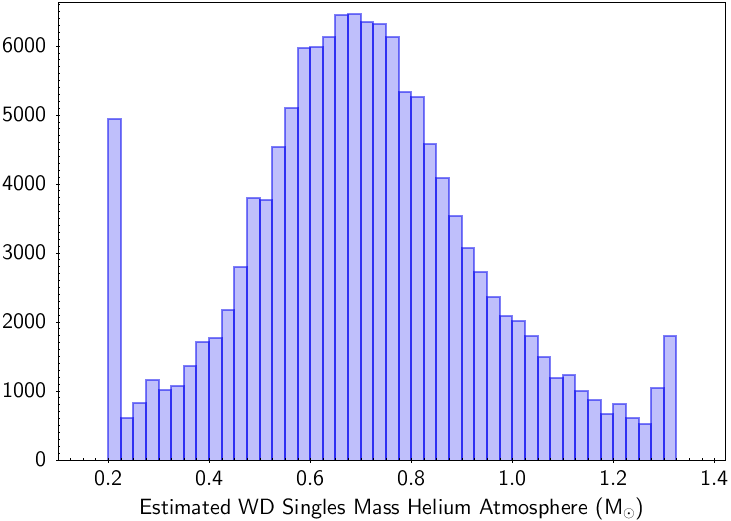}
    \caption{The estimated mass distribution of WDs identified as WD stars in Gaia filters assuming helium atmospheres (DB models). The sudden spikes on the tails of the distribution are likely WDNDs mixed into the single WD data and did not properly fit our models.}
    \label{fig:wd_dist_he}
\end{figure}

\begin{figure}
	\includegraphics[width=\columnwidth]{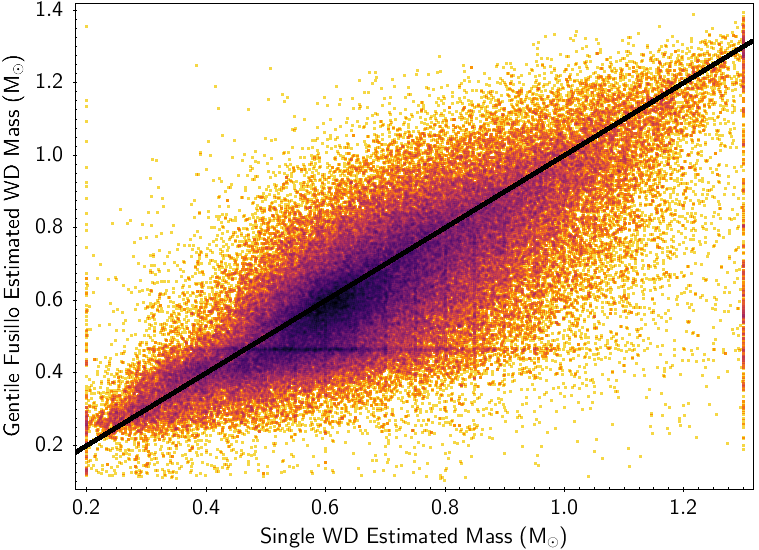}
    \caption{The estimated WD mass from \cite{2021MNRAS.508.3877G} WD catalogue compared to the estimated WD mass from our catalogue (DA models). The stars shown in the figure are a one-to-one match with 
    those in the \cite{2021MNRAS.508.3877G} catalogue and  
    contain only single WDs. The black line describes a one-to-one mass relationship between \cite{2021MNRAS.508.3877G} and our catalogue's estimated WD mass.}
    \label{fig:fusillo_vs_fitted}
\end{figure}

\begin{figure}
	\includegraphics[width=\columnwidth]{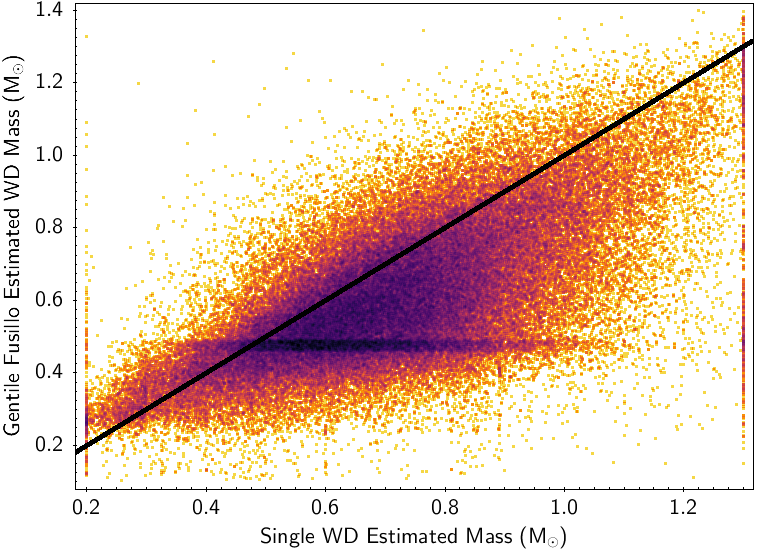}
    \caption{The estimated WD mass from \cite{2021MNRAS.508.3877G} WD catalogue compared to the estimated WD mass from our catalogue (DB models). The stars shown in the figure are a one-to-one match with 
    those in the \cite{2021MNRAS.508.3877G} catalogue and  
    contain only single WDs. The black line describes a one-to-one mass relationship between \cite{2021MNRAS.508.3877G} and our catalogue's estimated WD mass.}
    \label{fig:fusillo_vs_fitted_he}
\end{figure}

\begin{figure}

	\includegraphics[width=\columnwidth]{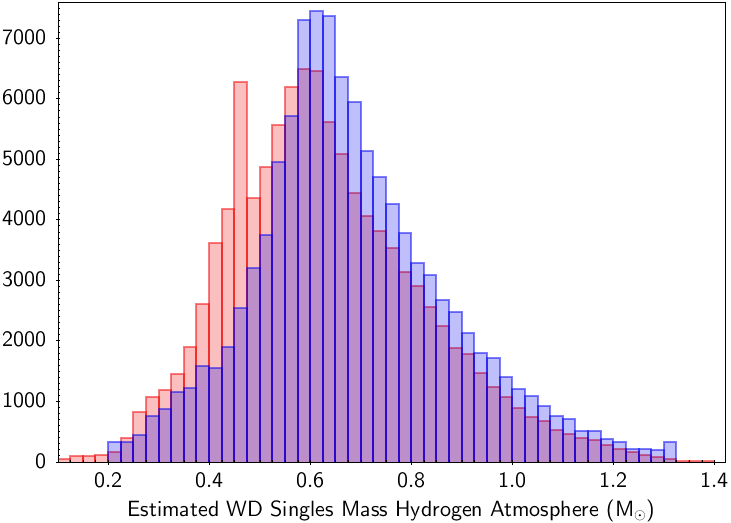}
    \caption{WD mass distribution of objects that only have entries in the \cite{2021MNRAS.508.3877G} catalogue (DA models). The blue represents our mass estimate and the orange represents the mass reported by \cite{2021MNRAS.508.3877G}. There is a clear shift to higher mass WDs in our estimated WD mass than those reported in \cite{2021MNRAS.508.3877G}.}
    \label{fig:fusillo_vs_fitted_hist}
\end{figure}

\begin{figure}

	\includegraphics[width=\columnwidth]{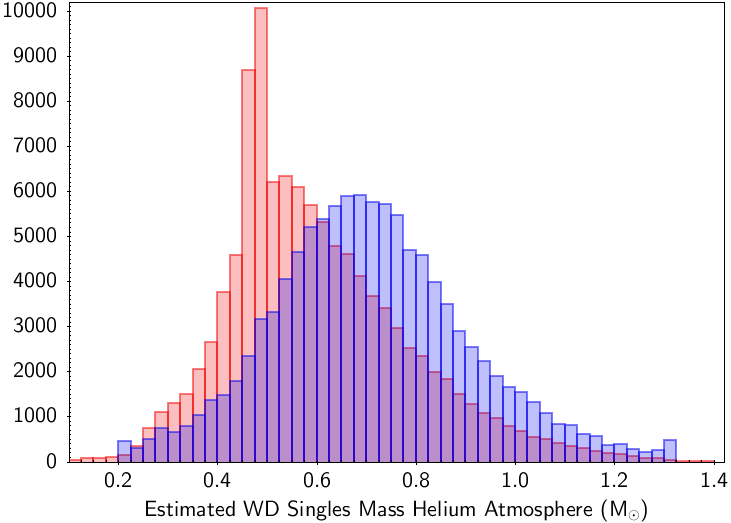}
    \caption{WD mass distribution of objects that only have entries in the \cite{2021MNRAS.508.3877G} catalogue (DB models). The blue represents our mass estimate and the orange represents the mass reported by \cite{2021MNRAS.508.3877G}. Similar to Figure \ref{fig:fusillo_vs_fitted_hist}, there is a clear shift to higher mass WDs in our estimated WD mass than those reported in \cite{2021MNRAS.508.3877G}.}
    \label{fig:fusillo_vs_fitted_hist_he}
\end{figure}

\begin{figure*}

    \includegraphics[width=0.49\linewidth]{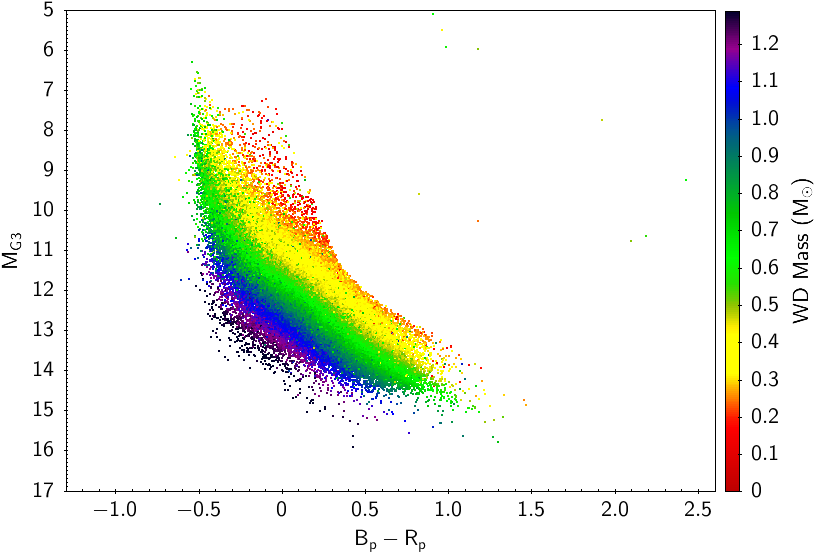} 
    \includegraphics[width=0.49\linewidth]{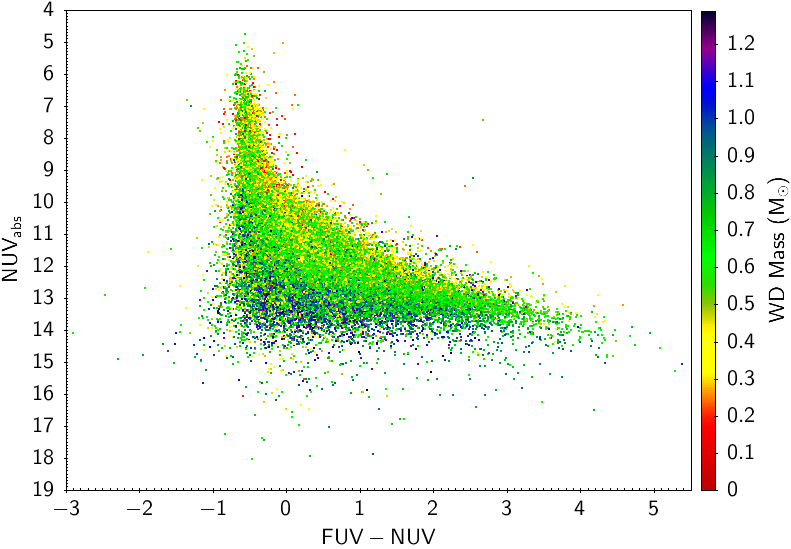}
    \includegraphics[width=0.49\linewidth]{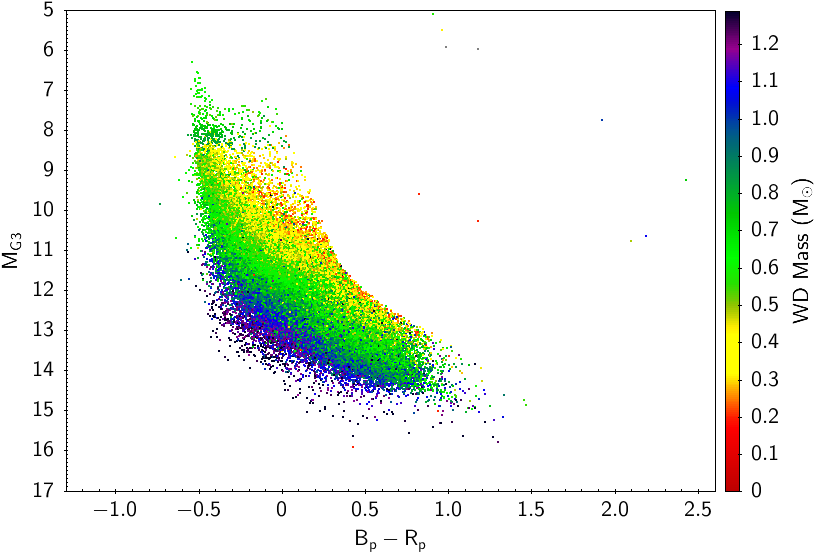}
    \includegraphics[width=0.49\linewidth]{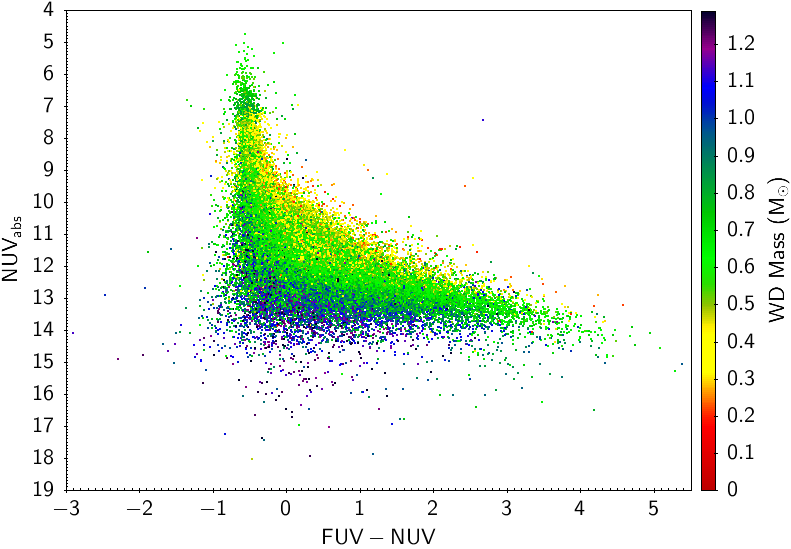}
    \caption{The four figures above depict the mass distribution of single WD stars in our catalogue that were previously identified in \cite{2021MNRAS.508.3877G} with hydrogen DA models. The top two figures depict the mass distribution reported by \cite{2021MNRAS.508.3877G} for both the visual magnitude in Gaia EDR3 on the left and the UV magnitude in GALEX on the right. The two bottom figures depict the mass distribution of WDs that we estimated considering the UV magnitude. The left figure depicts the visual magnitude in Gaia EDR3 on the left and the UV magnitude in GALEX on the right. 
    }
    \label{fig:mass_gaia_galax_est}

\end{figure*}

A significant portion of our catalogue contains single WDs. An attempt to estimate the mass, temperature and cooling age of the WD was performed using the photometric data and non-binary WD models. 
The mass distribution of these estimates, using DA model assumptions on the WD, can be observed in Figure \ref{fig:wd_dist} and using DB model assumptions in Figure \ref{fig:wd_dist_he}. 
The increase in objects at $0.2 \textrm{M}_\odot $ is likely caused by a fitting failure. This failure may have occurred due to the assumption that the WD is a single star, when in fact these WD systems may be WDND binary systems that were indistinguishable from those on the WD cooling sequence. The area used to define the single WD region is described by the blue area in the lower left corner of Figure \ref{fig:cmd}. One can observe a part of the WDND region is contained within the blue area, giving credibility to the idea that our single WD assumption is not accurate for all these systems. A similar problem may be occurring at $1.3 \textrm{M}_\odot$. To ensure our fitting function is accurate, we used a cross-match with the \cite{2021MNRAS.508.3877G} WD catalogue. This catalogue is primarily based on single WD systems gathered from Gaia EDR3 and estimates the WD mass in a similar $\chi^2$ minimization analysis. The matched entries were primarily single WDs and represent 111,996 entries in our catalogue. Figure \ref{fig:fusillo_vs_fitted} shows the distribution of DA WD mass estimates using GALEX and Gaia EDR3 data compared to the mass reported by \cite{2021MNRAS.508.3877G}. Figure \ref{fig:fusillo_vs_fitted_hist_he} shows the same distribution assuming DB WDs. The distribution of WD masses implies similar WD mass values to those found in \cite{2021MNRAS.508.3877G} with a few deviations. In Figure \ref{fig:fusillo_vs_fitted_hist}, the overall distribution is shifted towards somewhat higher mass WDs when compared to \cite{2021MNRAS.508.3877G}. This discrepancy between the masses in the two catalogues is likely caused by the inclusion of all available photometry from GALEX, SDSS and Gaia EDR3, where Gentile Fusillo \cite{2021MNRAS.508.3877G} only considers Gaia EDR3 data.  There is also an anomaly in the \cite{2021MNRAS.508.3877G} data that reports an unusually large number of WDs as having a mass of roughly 0.46 $\textrm{M}_\odot$. The source of this anomaly is unknown. We calculate the reduced $\chi^2$ between our WD mass estimates and \cite{2021MNRAS.508.3877G} WD mass estimates
\[
\chi^2_\textrm{\scriptsize red} = \frac{1}{N}\sum_i \left [ \frac{\left (M_\textrm{\scriptsize Fusillo, H}-M_\textrm{\scriptsize our, H} \right)^2}{{\sigma^2_{M,\textrm{\scriptsize Fusillo, H}}+\sigma^2_{M,\textrm{\scriptsize our, H}}}} \right ]
\]
over the WDs in both samples excluding the anomalous regions of $0.46 \textrm{M}_\odot <M_\textrm{\scriptsize Fusillo, H}<0.47 \textrm{M}_\odot $, $M_\textrm{\scriptsize our, H}<0.21 \textrm{M}_\odot $ and $M_\textrm{\scriptsize our, H}>1.29 \textrm{M}_\odot $ and obtain 2.5. The median value reported by our WD mass estimates excluding the anomalous regions is 0.66 and the median value reported by \cite{2021MNRAS.508.3877G} is 0.62, leading to a difference of 0.04. These values indicate that the WD mass estimates from our catalogue and Gentile-Fusillo catalogue generally agree within their respective uncertainties. 

\begin{figure*}

    \includegraphics[width=0.49\linewidth]{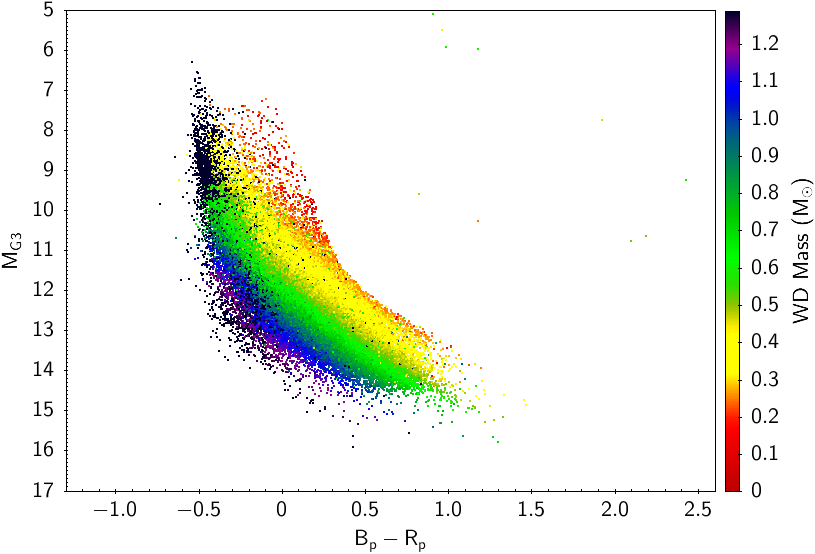} 
    \includegraphics[width=0.49\linewidth]{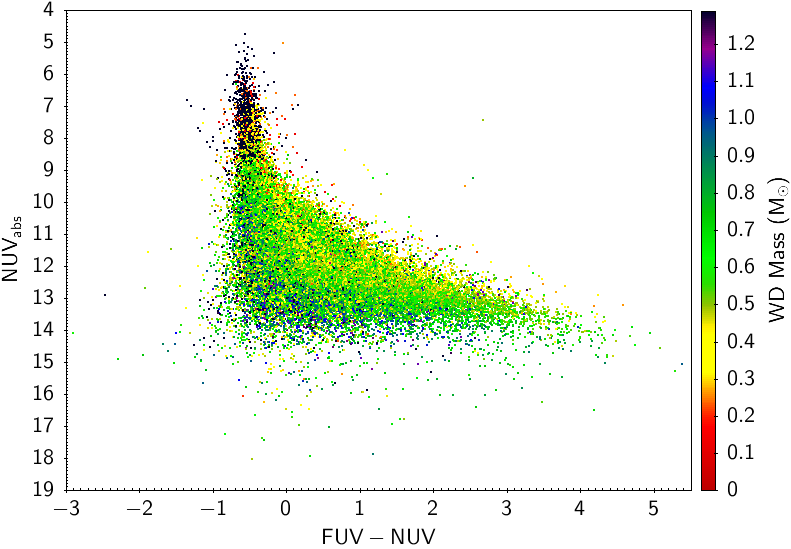}
    \includegraphics[width=0.49\linewidth]{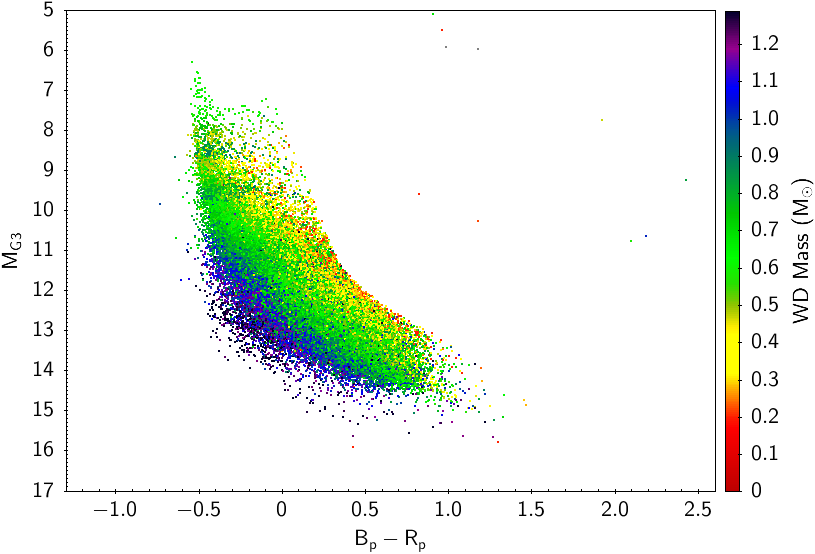}
    \includegraphics[width=0.49\linewidth]{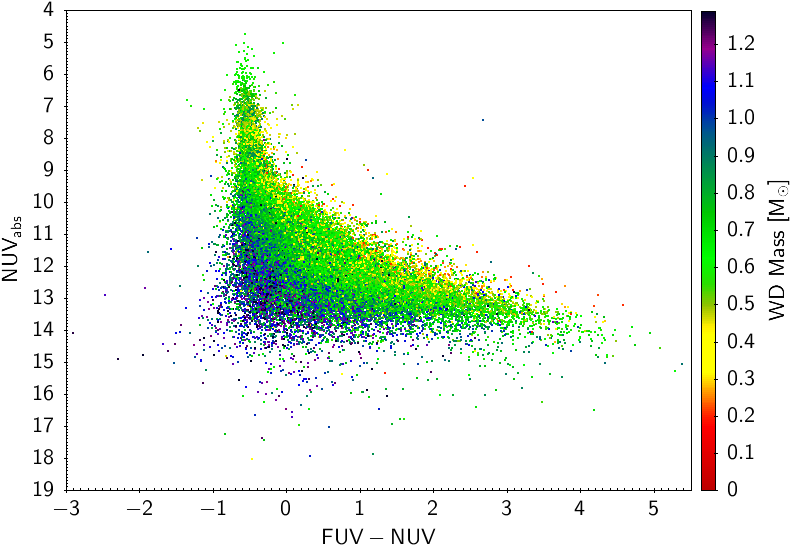}
    \caption{The four figures above depict the mass distribution of single WD stars in our catalogue that were previously identified in \cite{2021MNRAS.508.3877G} to have helium DB models. The top two figures depict the mass distribution reported by \cite{2021MNRAS.508.3877G} for both the visual magnitude in Gaia EDR3 on the left and the UV magnitude in GALEX on the right. The two bottom figures depict the mass distribution of WDs that we estimated considering all available photometric data. The left figure depicts the visual magnitude in Gaia EDR3 and the UV magnitude in GALEX on the right. 
    }
    \label{fig:mass_gaia_galax_est_he}

\end{figure*}

The WD DA estimated masses reported by \cite{2021MNRAS.508.3877G} are plotted onto both Gaia EDR3 CMDs and GALEX CMDs in the top two plots, alongside our estimates for DA WDs on the bottom two plots of Figure \ref{fig:mass_gaia_galax_est}. 
The bottom left plot in Figure \ref{fig:mass_gaia_galax_est} depicts single WDs with a broader distribution of WDs between 0.6-0.8 $\textrm{M}_\odot$ compared to the top left plot. In the top right plot, there is a far less homogeneous distribution in the UV for the reported values of WDs in \cite{2021MNRAS.508.3877G}. In particular, the high mass WD region has a high degree of scatter. However, in the bottom right plot of Figure \ref{fig:mass_gaia_galax_est} there is a spread of mass in the WDs with closer consistency to modeled WD masses. A similar pattern is observed with the DB WDs in Figure \ref{fig:mass_gaia_galax_est_he}. It is uncertain how accurate the GALEX UV data is when predicting the WD masses, but Figure \ref{fig:mass_gaia_galax_est} indicates that there is a significant change when including UV data. 

By construction, each object in our catalogue has a measured UV flux either in the NUV or the FUV.  This information clearly influences the masses determined through the fitting process as our fitted masses better correlate with a star's position in the UV CMD than the masses from \cite{2021MNRAS.508.3877G} as shown in Fig.~\ref{fig:mass_gaia_galax_est} and~\ref{fig:mass_gaia_galax_est_he}. The additional UV data provides both an opportunity in that the UV fluxes depend strongly on the temperature of the white dwarfs and a potential pitfall in that the observed UV fluxes typically have larger statistical and systematic uncertainties than the Gaia DR3 fluxes.   As Fig.~\ref{fig:fusillo_vs_fitted} through~\ref{fig:fusillo_vs_fitted_hist_he}, our fitted masses (whether using DA or DB models) are systematically slightly larger than those obtained by \cite{2021MNRAS.508.3877G}.  Adding the UV data to the visual data from Gaia used by \cite{2021MNRAS.508.3877G} may yield more accurate temperature estimates and, therefore, more accurate mass estimates. However, there is a bias in that the GALEX flux measurements have larger uncertainties than the Gaia measurements. These uncertainties, when combined with the flux distribution of WDs (fainter WDs), will systematically cause the UV fluxes to be overestimated, thereby the temperatures also to be overestimated and finally also the masses to be overestimated.  

\subsection{White Dwarfs in Binaries}

WDND WD mass estimates were less successful than single WD mass estimates. 
A large portion of the WD mass estimates for WDNS systems were approximately $0.2 \textrm{M}_\odot$, $0.3 \textrm{M}_\odot$ and $1.3 \textrm{M}_\odot$. The large fraction of WD estimates at $0.3 \textrm{M}_\odot$ is related to the initial guess used in the $\chi^2$ fitter for the WD mass which was given as $0.3 \textrm{M}_\odot$. All three values are failed WD mass estimates where the fitter was unable to gather reliable mass estimates. 
In areas outside these three values, the masses returned by our fitting were uniform over the allowed range in the fitting. The WD model takes up any UV excess in the data relative to the MS model, so discrepancies in the UV data, the MS or giant models, the WD models or all three yield inaccurate WD mass estimates.
As stated in Section 2.3, the models used to calculate the GALEX NUV output of MS stars fitted with Gaia EDR3 data exhibited considerably more NUV emission than observed. Consequently, it is likely that the poor reliability of many of our estimates is related to a lack of suitable models for the GALEX FUV and NUV magnitudes of the stars. We cannot estimate the mass of WDs in WDNDs with photometry alone and spectra are required to have certainty of WDND WD mass estimates. 

\begin{figure}
	\includegraphics[width=\columnwidth]{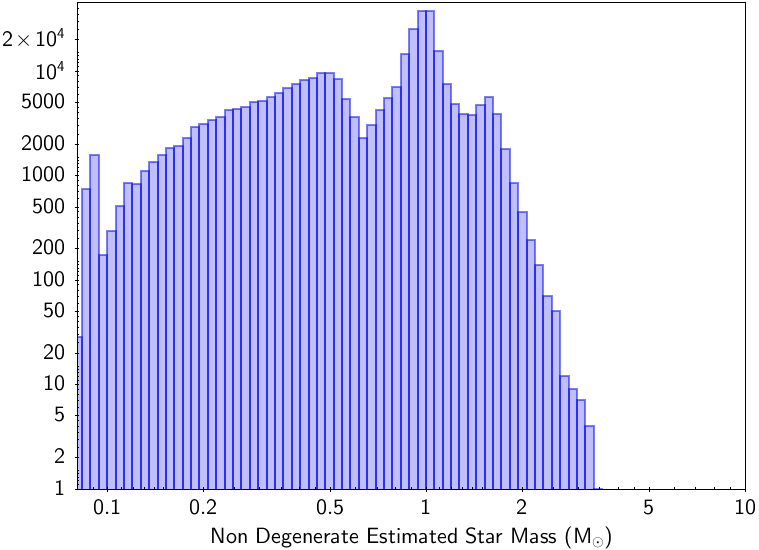}
    \caption{The distribution of mass estimates of non-degenerate stars with a WD companion contributing to the UV of the system. For objects above 0.9 $\textrm{M}_\odot$, the distribution of estimated stellar mass becomes unreliable. This is due to the assumed WD in the system dominating the UV, forcing the non-degenerate mass estimate to be smaller than what can be seen in the optical. There is also a discontinuity in our fitting due to the difficulty with the isochrones for giant stars that can be observed as the peaks in Figure \ref{fig:starhorse_vs_fitmass}.}
    \label{fig:sec_m_hist}
\end{figure}

\begin{figure}
	\includegraphics[width=\columnwidth]{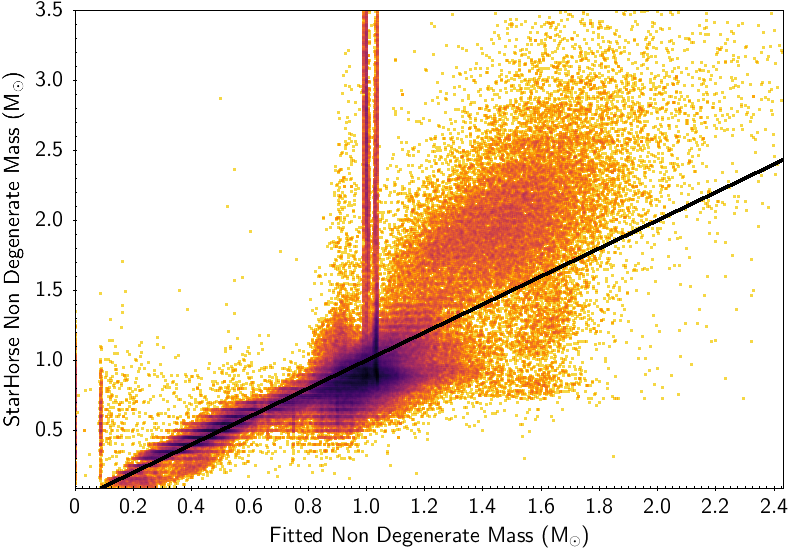}
    \caption{Comparison of the difference in our fitted secondary star masses and masses in StarHorse. The black line describes the situation where the mass we have fitted perfectly matches the StarHorse mass. Objects below the black line are stars where we fit to higher masses than StarHorse and above the black line are stars where we fit a smaller mass than StarHorse. The spiked lines at 1$\textrm{M}_\odot$ and 1.04$\textrm{M}_\odot$ are known issues with the isochrones used being incapable of correctly estimating the mass of evolved stars.}
    \label{fig:starhorse_vs_fitmass}
\end{figure}

\begin{figure}
	\includegraphics[width=\columnwidth]{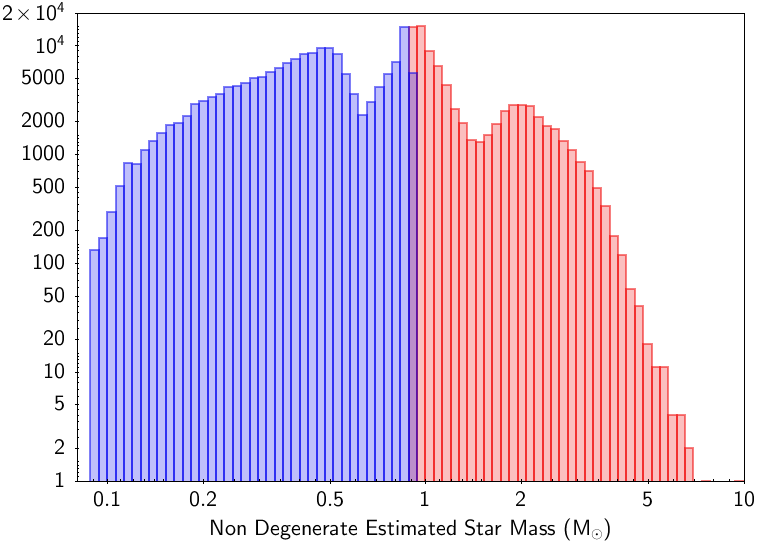}
    \caption{This distribution represents our estimated non-degenerate mass in the blue with StarHorse estimated mass in the red from 0.09 $\textrm{M}_\odot$ to 9.53 $\textrm{M}_\odot$. A cutoff between the two distributions is performed at 0.9 $\textrm{M}_\odot$ where we believe that StarHorse MS mass estimates will be less reliable below 0.9 $\textrm{M}_\odot$ due to the presence of a WD in the system.}
    \label{fig:sec_m_hist_2}
\end{figure}

These WDND candidates are MS or evolved stars whose photometry exhibits a UV excess similar to WD stars. 
If these are WDND systems, we expect the WD UV magnitude to either dominate the UV light observed or have an additional contribution of UV flux from the non-degenerate star. This would move the object off the WD sequence described below the black line in the left plot of Figure \ref{fig:image2} and toward the top right of the figure.
Below the WD region depicted in the left plot of Figure \ref{fig:image2}, the catalogue does not likely contain high mass WDs. Instead, there is a possibility that these are low flux readings from MS stars or flaring M-dwarf stars. This is consistent with SDSS spectra classifications shown in Figure \ref{fig:sdss_spectra} where M-dwarf stars are identified in our catalogue based on their UV data, not their optical data. This suggests that a certain fraction of the catalogue contains M-dwarf stars with excess UV. 
A $\chi^2$ minimization fit was performed to attempt to identify the masses of these non-degenerate systems assuming a potential WD was contributing to the UV flux. The distribution of non-degenerate mass estimates can be seen in Figure \ref{fig:sec_m_hist}. A cross-match with the StarHorse MS catalogue \citep{2021arXiv211101860A} was performed to supplement our non-degenerate mass estimates. Of the 309,584 entries, 193,221 objects matched to StarHorse and reported a non-degenerate mass estimate. Figure \ref{fig:starhorse_vs_fitmass} describes the difference between our non-degenerate mass estimates and StarHorse's reported MS mass. The black line in the figure represents a one-to-one relationship. The stars above the black line are stars where StarHorse reported masses higher than our own estimates and below the black line represents stars where StarHorse reported a lower mass than our estimates. From Figure \ref{fig:starhorse_vs_fitmass}, there exists a one-to-one correlation between our estimated masses and the StarHorse reported masses up until 0.9 $\textrm{M}_\odot$. A discontinuity exists at 1.04 $\textrm{M}_\odot$. This is a known degeneracy in our fitting caused by the upper mass limit of one of the isochrones used. 
In order to develop a complete look at the non-degenerate mass distribution of the catalogue, we utilize a mass cut. 
For stars with masses greater than $0.9 \textrm{M}_\odot$, StarHorse is likely the more accurate estimate for the non-degenerate mass. This is due to the uncertainty in the UV flux of the models we used, 
whereas for stars less massive than $0.9 \textrm{M}_\odot$, we assume that StarHorse will incorrectly estimate the mass of the non-degenerate star if there is a WD in the system. This is due to the non-degenerate star being faint enough to allow for the WD to contribute significantly to the visual flux of the system. 
Figure \ref{fig:sec_m_hist_2} depicts the distribution of non-degenerate star masses using our mass estimates for non-degenerate stars with a mass less than $0.9 \textrm{M}_\odot$ (shown in blue) and StarHorse mass estimates for stars with a mass greater than $0.9 \textrm{M}_\odot$ (shown in orange). The slopes of the non-degenerate mass distribution of Figure \ref{fig:sec_m_hist_2} change considerably. 
On the low mass end, the slope incrementally increases, then flattens out to an exponent of 1.4. A dearth of objects exists at $0.65 \textrm{M}_\odot$. 
On the high mass end, the distribution displays another sudden lack of objects at around $1.5 \textrm{M}_\odot$. After the dip, the slope of the distribution is a $-4$ exponent before becoming a slope of $-8$ exponent. We calculate that for binary systems where one star has evolved into a WD, the initial mass function (IMF) should tend toward a slope of $-7.5$ exponent. \cite{2017ApJS..230...15M} reported that in recent observations of binary systems IMFs with smaller mass ratios (0.1-0.3), the power-law of the mass ratio distribution flattens towards shallower slopes. 

\begin{figure}
	\includegraphics[width=\columnwidth]{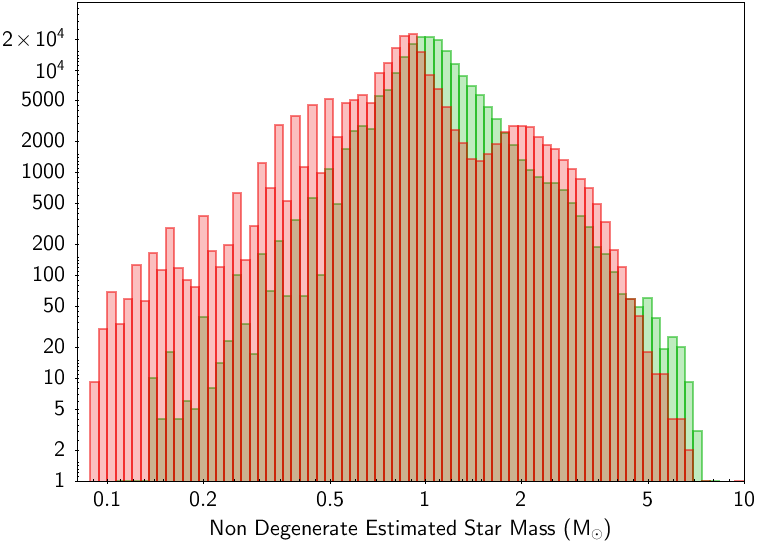}
    \caption{Comparison of the StarHorse mass distribution of stars found in our candidate WDND systems correlate to the red distribution. A general distribution of StarHorse main sequence masses are observed in green. Each distribution contains the same number of stars. For objects below 0.9 $\textrm{M}_\odot$, the distribution of estimated stellar mass becomes unreliable. This is due to the StarHorse catalogue assuming that the system is a single star system. }
    \label{fig:sec_m_hist3}
\end{figure}


\subsection{SDSS Spectroscopy}

\begin{figure}
	\includegraphics[width=\columnwidth]{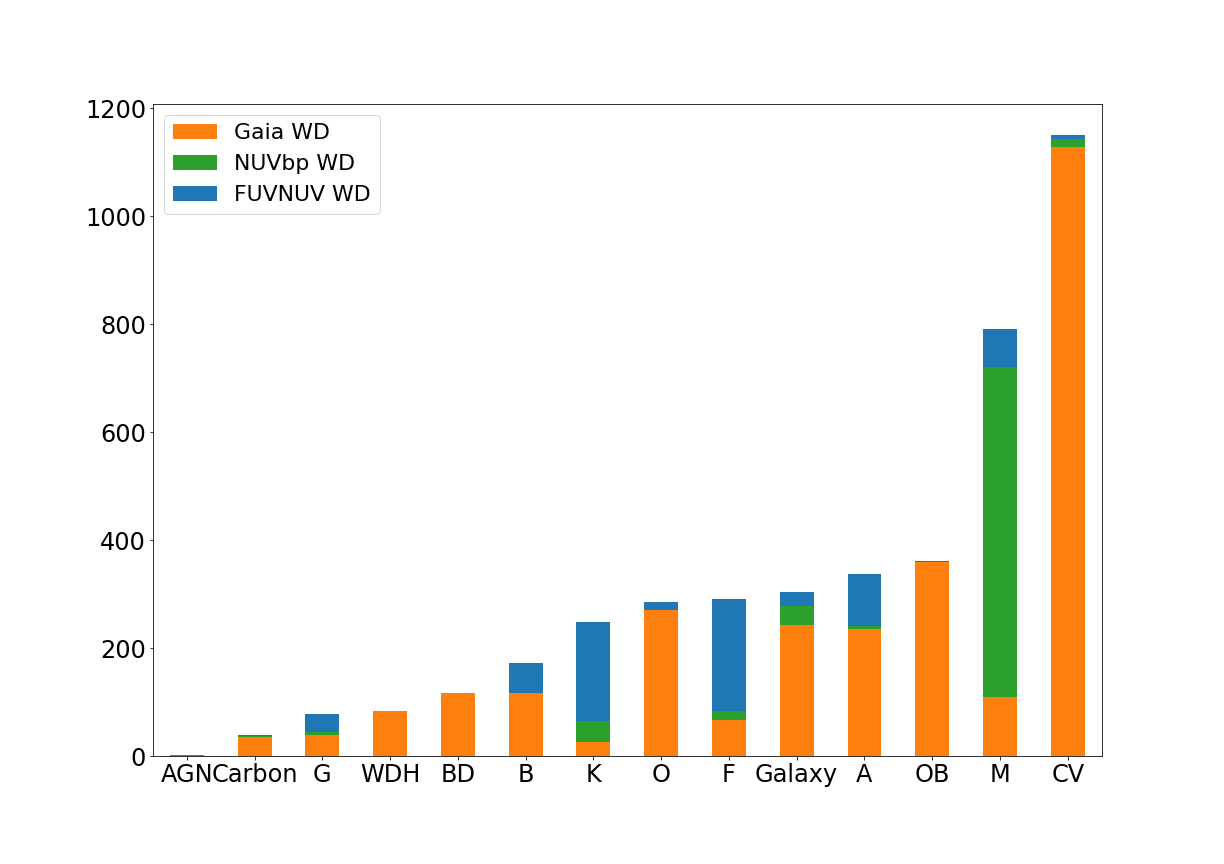}
    \caption{The distribution of the 11,607 different SDSS DR12 spectral classifications. There are 6,432 WD spectral classifications that are not listed in the figure above. The orange depicts the distribution of SDSS classifications that appear to be WDs in Gaia EDR3 photometry, the green depicts the distribution of objects with SDSS classification that appear to be WDs in the GALEX NUV and Gaia EDR3 Bp region and the blue depicts the SDSS classification that appear to be WDs in the GALEX photometry alone. The distribution is dominated by over 6,000 WD identifications which are not depicted in histogram.}
    \label{fig:sdss_spectra}
\end{figure}

\begin{figure}
	\includegraphics[width=\columnwidth]{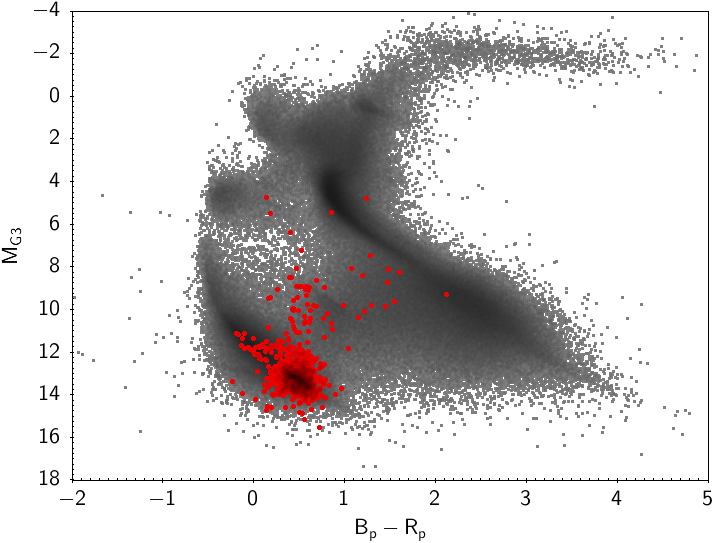}
    \caption{The depiction of CV objects in our catalogue CMD in Gaia EDR3 magnituddes. All objects in the catalogue are marked in grey. The red points mark the objects that were classified by SDSS DR12 as CVs.}
    \label{fig:gaia_cv}
\end{figure}

All entries with available SDSS photometry and spectral data have been included in the catalogue. 
A total of 11,607 entries have SDSS spectral classifications. Figure \ref{fig:sdss_spectra} depicts the distribution of the various spectral types identified by SDSS. The distribution of SDSS classifications is heavily biased to WD identifications with 6,432 of the 11,607 objects being single WDs. 
Of the remaining entries, 1,151 were listed as cataclysmic variables (CVs). These objects can be seen in the Gaia EDR3 CMD in Figure \ref{fig:gaia_cv}. There is a clear concentration of objects identified as CVs on the WD sequence. Upon investigating their spectra, a large portion of these objects do not appear to have clear emission lines. The SDSS DR12 website provides the following caveat: ``Although sub-classifications are provided, they are not optimized for accuracy. In particular, the CV star templates have more degrees of freedom than other stellar templates, which can result in unphysical solutions where negative PCA components of the CV templates can fit absorption features of White Dwarfs" \citep{WinNT}. It is unlikely that the 1,151 listed CVs are accurately identified; therefore, a significant portion of CVs listed are likely single WDs. 
A total of 270 or 2.3 percent of the SDSS entries that matched with our catalog were listed as galaxies. Using the SDSS optical spectrum archive, we examined the top ten galaxy entries with the highest signal-to-noise ratio to further understand if there was a large contaminant of galaxies in our data. Of these ten entries, nine were misidentified. Spectra with clear WD features often have their hydrogen absorption lines misidentified as oxygen II emission lines, giving several entries a consistent redshift of roughly $0.7$. There is also misidentification of M-dwarf stars as galaxies due to the presence of $H_\alpha$ and $H_\beta$ in the spectrum. The presence of the hydrogen spectral lines in the M-dwarf spectra likely causes the spectral identifier to assume the system contains many stars and is, thus, a galaxy. It is also likely that the source of these hydrogen spectral lines could be coming from a WD in the system and not a star-rich galaxy. Furthermore, several entries had a redshift close to zero or were actually negative, meaning they are likely misidentified stars, not galaxies. After examining the spectra of the 270 listed galaxies, it seems that a large fraction of them are misidentified sources. It is worth noting that these 270 objects make up only a tiny fraction of the 791,977 galaxy identifications in the SDSS DR12 catalogue \citep{2015ApJS..219...12A}.

\section{Conclusions}

The aim of the GALEX-Gaia-EDR3 catalogue is to identify and characterise candidate white-dwarf-non-degenerate binary stars to be used for additional study or for follow-up observation. The GALEX-Gaia-EDR3 catalogue was constructed from GALEX and Gaia EDR3 catalogues containing 332,111 candidate white dwarf binary systems and 111,996 candidate single white dwarfs. When available, data from SDSS DR12, Pan-STARRS DR1, StarHorse MS catalogue 
and Gentile-Fusillo Gaia EDR3 white dwarf catalogue augment the catalogue to provide supplementary data. Estimates for the white dwarf's mass, age, effective temperature and the non-degenerate secondary mass were determined through fitting the photometry of objects in our catalogue with modeled white dwarf cooling sequences from \cite{2020ApJ...901...93B}. Upon comparing our results with the 111,996 candidate single white dwarfs with the Gentile-Fusillo Gaia EDR3 white dwarf catalogue, we found that our estimates for the single white dwarf's mass tended to be skewed slightly towards higher-mass white dwarfs, but generally agree within their respective uncertainties. Mass estimates for the white dwarfs in binary systems have a high degree of inaccuracy due to the presence of the non-degenerate star significantly contributing to the photometry of the system. 

\section*{Acknowledgements}

This research has made use of the VizieR catalogue access tool, CDS,
 Strasbourg, France (DOI : 10.26093/cds/vizier). The original description 
 of the VizieR service was published in 2000, A\&AS 143, 23

This work has made use of data from the European Space Agency (ESA) mission
{\it Gaia} (\url{https://www.cosmos.esa.int/gaia}), processed by the {\it Gaia}
Data Processing and Analysis Consortium (DPAC, \url{https://www.cosmos.esa.int/web/gaia/dpac/consortium}). Funding for the DPAC has been provided by national institutions, in particular the institutions
participating in the {\it Gaia} Multilateral Agreement.

Funding for SDSS-III has been provided by the Alfred P. Sloan Foundation, the Participating Institutions, the National Science Foundation, and the U.S. Department of Energy Office of Science. The SDSS-III web site is \url{http://www.sdss3.org/}.

SDSS-III is managed by the Astrophysical Research Consortium for the Participating Institutions of the SDSS-III Collaboration including the University of Arizona, the Brazilian Participation Group, Brookhaven National Laboratory, Carnegie Mellon University, University of Florida, the French Participation Group, the German Participation Group, Harvard University, the Instituto de Astrofisica de Canarias, the Michigan State/Notre Dame/JINA Participation Group, Johns Hopkins University, Lawrence Berkeley National Laboratory, Max Planck Institute for Astrophysics, Max Planck Institute for Extraterrestrial Physics, New Mexico State University, New York University, Ohio State University, Pennsylvania State University, University of Portsmouth, Princeton University, the Spanish Participation Group, University of Tokyo, University of Utah, Vanderbilt University, University of Virginia, University of Washington, and Yale University. 

\section*{Data Availability}

The catalogue presented in this work can be downloaded \href{https://doi.org/10.5281/zenodo.8145200}{HERE} or at the url: \url{https://doi.org/10.5281/zenodo.8145200}. The catalogue will also be made available via the VizieR Service for Astronomical Catalogues. The original data in this article is publicly available from the relevant survey archives.



\bibliographystyle{mnras}
\bibliography{main}




\appendix

\section{Column Descriptions}

ra\_epoch2000, Right Ascension of the object taken from Gaia at Epoch2000

dec\_epoch2000, Declination of the object taken from Gaia at Epoch2000

ra\_epoch2000\_error, Error of the Right Ascension of the object taken from Gaia at Epoch2000 

dec\_epoch2000\_error, Error of the Declination of the object taken from Gaia at Epoch2000

ra\_dec\_epoch2000\_corr, Correlation between right ascension and declination

source\_id, a Unique source identifier within a Gaia data release

name, GALEX object name

objid, GALEX identifier in JHHMMSS.s+DDMMSS format

Cat, Survey type (Always AIS-all-sky imaging survey)

FUVexp, FUV exposure time. 

NUVexp, NUV exposure time

GLON, Galactic Longitude

GLAT, Galactic Latitude

tile, Title number

img, Image number

sv, Number of subsists if exposure was divided

r.fov, Distance of source from center of the field

Obs, Observation type

b, Bands: 1=NUV, 2=FUV, 3 both

E(B-V), Galactic reddening inferred from 100um dust emission maps (E\_bv)

Sp?, 1 if a spectrum exists (istherespectrum)

chkf, Astrometry check type

FUVmag, Galex FUV Calibrated magnitude in AB system

e\_FUVmag, FUV mag error

NUVmag, Galex NUV calibrated magnitude in AB system

e\_NUVmag, NUV mag error

FUV.a, FUV Kron-like elliptical aperture magnitude

e\_FUV.a, FUV RMS error for AUTO magnitude 

NUV.a, NUV Korn-like elliptical aperture

e\_NUV.a, NUV RMS error for AUTO magnitude

FUV.4, FUV mag aperture 8 pix diameter

e\_FUV.4, FUV mag aperture error (8 pix diameter)

NUV.4, NUV mag aperture (8 pix diameter)

e\_NUV.4, NUV mag aperture error (8 pix diameter)

FUV.6, FUV mag aperture (17 pix diameter)

e\_FUV.6, FUV mag aperture error (17 pix diameter)

NUV.6, NUV mag aperture (17 pix diameter)

e\_NUV.6, NUV mag aperture error (17 pix diameter)

Fafl, FUV artifact Flag (logical OR near source)

Nafl, NUV artifact Flag (logical OR near source)

Fexf, Extraction flags for FUV band

Nexf, Extraction flags for NUV band

Fflux, FUV calibrated Flux

e\_Fflux, FUV flux error

Nflux, NUV calibrated flux

e\_Nflux, NUV flux error

FXpos, Source position in FUV image along x

FYpos, Source position in UV image along y

NXpos, Source position in NUV image along x

NYpos, Source position in NUV image along y

Fima, Source FWHM in FUV assuming a gaussian core

Nima, Source FWHM in NUV assuming a gaussian core

Fr, Source FWHM  in FUV assuming a gaussian core

Nr, Source FWHM in NUV assuming a gaussian core

nS/G, Star/galaxy classifier from NUV 

fS/G, Star/galaxy classifier from FUV

nell, 1-b/a in NUV 

fell, 1-b/a in FUV

nPA, Position angle in NUV

e\_nPA, Position angle error in NUV

fPA, Position angle in FUV

e\_fPA, Position angle error in FUV

Fnr, FUV FWHM IMAGE value from -fd-ncat.fits

F3r, FUV flux radius 

Nar, Kron apertures in units of a or b for NUV

Narms, Profile RMS along major axis for NUV

Nbrms, Profile RMS along minor axis for NUV

Far, Kron apertures in units of a or b for FUV

Farms, Profile RMS along major axis for FUV 

Fbrms, Profile RMS along minor axis for FUV

w\_NUV, NUV effective exposure (flat-field response value) at the source position

w\_FUV, FUV effective exposure

Prob, Probability of the FUVxNUV match

Sep, Separation between FUV and NUV position of the source in the same observation

Nerr, Position error of the source in the NUV image

Ferr, Position error of the source in the FUV image

Ierr, Inter-band position error

Nperr, NUV poisson position error

Fperr, FUV poisson position error

CV, Whether the source comes from a coda or visit. (corv)

G, Neighbours rank

N, Number of sources within 2.5’’, if this is a primary.

primid, objid of the closest primary, based on distance criterion

groupid, Objid’s of all sources (AIS) within 2.5’’, concatenated by “+”

Gd, as for “G” (grank), but based on distance criterion (granddist)

Nd, as for “N” (ngrank), but based on distance criterion

OName, Extended object if not:”N”

parallax, Absolute stellar parallax

parallax\_error, standard error of parallax

pm, Total proper motion (pm)

pmra, proper motion in right ascension direction

pmra\_error, standard error of proper motion in right ascension direction

pmdec, proper motion in declination direction

pmdec\_error, standard error of proper motion in declination direction

astronomic\_n\_good\_obs\_al, number of good observations AL

astronomic\_gof\_al, Goodness of fit statistic of model wrt along-scan observations

astrometric\_chi2\_al, AL chi-squared value

astrometric\_excess\_noise, Excess noise of the source

astrometric\_excess\_noise\_sig, Significance of excess noise

astrometric\_params\_solved, Which parameters have been solved for

pseudocolour, astrometrically determined pseduocolour of the source

pseudocolour\_error, standard error of pseudo color of source

visibility\_periods\_used, Number of visibility periods used in astronomic solution

ruwe, Gaia dr2 re-normalised unit-weight error \citep{2018A&A...616A...2L}

duplicated\_source, Source with duplicate sources

phot\_g\_mean\_flux, Mean G Flux

phot\_g\_mean\_flux\_error, Mean G Flux error
             
phot\_g\_mean\_mag, mean G magnitude

phot\_g\_mean\_mag\_error, Mean G magnitude error

phot\_bp\_mean\_flux, Integrated mean flux in BP band

phot\_bp\_mean\_flux\_error, Error of the integrated BP mean flux

phot\_bp\_mean\_mag, Integrated BP mean magnitude (Vega)

phot\_bp\_mean\_mag\_error, error of the integrated BP mean magnitude

phot\_rp\_mean\_flux, Integrated RP mean flux

phot\_rp\_mean\_mag, Integrated RP mean magnitude (Vega)
             
phot\_rp\_mean\_mag\_error, error on integrated RP mean magnitude

phot\_bp\_rp\_excess\_factor, BP/RP excess factor

bp\_rp, BP-RP Gaia colour

phot\_g\_mean\_mag\_corrected, Calibration corrected G magnitude (added by CDS)

phot\_g\_mean\_mag\_error\_corrected, Standard error of calibration corrected G magnitude (added by CDS)

phot\_g\_mean\_flux\_corrected, Calibration corrected G-band mean flux (added by CDS)

phot\_bp\_rp\_excess\_factor\_corrected, Calibration corrected BP/RP excess factor (added by CDS)

dr2\_radial\_velocity, Radial velocity from Gaia DR2

dr2\_radial\_velocity\_error, Radial velocity error from Gaia DR2

dr2\_rv\_nb\_transits, Number of transits used to compute radial velocity in Gaia DR2

dr2\_rv\_template\_teff, Teff of the template used to compute radial velocity in Gaia DR2

dr2\_rv\_template\_logg, logg of the template used to compute radial velocity in Gaia DR2

urat1, URAT1 cross-id name (urat1)
             
objID\_psdr1, Unique object identifier for Pan-STARRS

f\_objID\_psdr1, Information flag bitmask indicating details of the photometry. 

Qual\_psdr1, Subset of objInfoFlag denoting whether this object is real or likely false positive. 

Epoch\_psdr1, Modified Julian date of the mean Epoch

Ns\_psdr1, Number of stack detection

gmag\_psdr1, Mean PSF AB magnitude from g filter (4866 angstroms) detection

e\_gmag\_psdr1, error in gmag

gKmag\_psdr1, mean Kron AB magnitude from g filter detection 

e\_gKmag\_psdr1, error in gKmag 

gFlags\_psdr1, Information flag bitmask for mean object from g filter detection
             
rmag\_psdr1, Mean PSF AB magnitude from r filter (6215 angstroms) detection. 

e\_rmag\_psdr1, error in rmag

rKmag\_psdr1, Mean Kron AB magnitude from r filter detection

e\_rKmag\_psdr1, Error in rKmag

rFlags\_psdr1, Information flag bitmask fro mean object from r filter detection

imag\_psdr1, mean PSF AB magnitude from i filer (7545 angstroms) detection. 

e\_imag\_psdr1, error in imag
             
iKmag\_psdr1, Mean Kron AB magnitude from i filter detection

e\_iKmag\_psdr1, error in iKmag 

iFlags\_psdr1, Information flag bitmask fro mean object from i filter detection

zmag\_psdr1, Mean PSF AB magnitude from z filter (8679 angstroms) detection

e\_zmag\_psdr1, error in zmag

zKmag\_psdr1, Mean Kron AB magnitude from z filter detection
             
e\_zKmag\_psdr1, Error in zKmag

zFlags\_psdr1, Information flag bitmask for mean object from z filter detection

ymag\_psdr1, Mean PSF AB magnitude from y filter (9633 angstroms) detection

e\_ymag\_psdr1, Error in ymag

yKmag\_psdr1, Mean Kron AB magnitude from y filter detection

e\_yKmag\_psdr1, error in yKmag

yFlags\_psdr1, Information flag bitmask for mean object from y filter detection

objID\_sdssdr12, SDSS unique object identifier

mode\_sdssdr12, 1: primary  (469,053,874 sources), 2: secondary (324,960,094 sources)

q\_mode\_sdssdr12, '+' indicates clean photometry (310,190,812 sources with mode 1+) (clean)

class\_sdssdr12, Type of object (3=galaxy, 6=star)

SDSS12\_sdssdr12, SDSS-DR12 name, based on J2000 position
             
m\_SDSS12\_sdssdr12, The asterisk indicates that 2 different SDSS objects share the same SDSS12 name

flags\_sdssdr12, Photo Object Attribute flags 

ObsDate\_sdssdr12, Mean Observation date

Q\_sdssdr12, Quality of the observation

umag\_sdssdr12, model magnitude in u filter, AB scale
             
e\_umag\_sdssdr12, mean error on umag

gmag\_sdssdr12, model magnitude in g filter, AB scale

e\_gmag\_sdssdr12, Mean Error on gmag

rmag\_sdssdr12, Model magnitude in r filter, AB scale

e\_rmag\_sdssdr12, Mean error on rmag
             
imag\_sdssdr12, Model magnitude in i filter, AB scale

e\_imag\_sdssdr12, Mean error on imag

zmag\_sdssdr12, Model magnitude in z filter, AB scale

e\_zmag\_sdssdr12, Mean error on zmag

zsp\_sdssdr12, Spectroscopic redshift (when SpObjID>0)
             
e\_zsp\_sdssdr12, Mean error on zsp

f\_zsp\_sdssdr12, “zwarning” flag

zph\_sdssdr12, Photometric redshift; estimated by robust fit to nearest neighbors in a reference set. 

e\_zph\_sdssdr12, estimated error of the photometric redshift

avg\_zph\_sdssdr12, Average redshift of the nearest neighbors; if significantly different from mph this might be a better estimate than zph. 

pmRA\_sdssdr12, Proper motion along Right Ascension

e\_pmRA\_sdssdr12, Mean error on pmRA

pmDE\_sdssdr12, Proper motion along Declination

e\_pmDE\_sdssdr12, mean error on pmDE

SpObjID\_sdssdr12, Pointer to the spectrum of object or 0

spType\_sdssdr12, Source type (sourceType)

spCl\_sdssdr12, Spectroscopic class: GALAXY, QSO, STAR

subClass\_sdssdr12, spectroscopic subclass

dist05, StarHorse distance, 5th percentile

dist16, StarHorse distance, 16th percentile

dist50, StarHorse distance, 50th percentile

dist84, StarHorse distance, 85th percentile

dist95, StarHorse distance, 95th percentile
             
av05, StarHorse line-of-sight extinction at lambda = 5420 angstrom, 5th percentile

av16, StarHorse line-of-sight extinction at lambda = 5420 angstrom, 16th percentile

av50, StarHorse line-of-sight extinction at lambda = 5420 angstrom, 50th percentile

av84, StarHorse line-of-sight extinction at lambda = 5420 angstrom, 84th percentile

av95, StarHorse line-of-sight extinction at lambda = 5420 angstrom, 95th percentile

teff16, StarHorse effective temperature, 16th percentile

teff50, StarHorse effective temperature, 50th percentile

teff84, StarHorse effective temperature, 84th percentile

logg16, StarHorse Surface Gravity, 16th Percentile

logg50, StarHorse Surface Gravity, 50th Percentile

logg84, StarHorse Surface Gravity, 84th Percentile

met16, StarHorse metallicity, 16th percentile
             
met50, StarHorse metallicity, 50th percentile

met84, StarHorse metallicity, 84th percentile

mass16, StarHorse Stellar Mass, 16th Percentile

mass50, StarHorse Stellar Mass, 50th Percentile

mass84, StarHorse Stellar Mass, 84th Percentile

‘fidelity', Gaia EDR3 astrometric fidelity flag (Rybizki et al., 2021, MNRAS, in prep.)

TeffH, Effective temperature from fitting the dereddened G, BP, and RP absolute fluxes with pure-H model atmospheres (teff\_H), from \cite{2021MNRAS.508.3877G}

e\_TeffH, Uncertainty of teff\_H (eteff\_H), from \cite{2021MNRAS.508.3877G}

loggH, Log of surface gravity from the dereddened G, BP, and RP absolute fluxes with pure-H model atmospheres (logg\_H) from \cite{2021MNRAS.508.3877G}

e\_loggH, Uncertainty on logg\_H (elogg\_H), from \cite{2021MNRAS.508.3877G}

MassH, Stellar mass resulting from the adopted mass-radius relation and best fit parameters (mass\_H), from \cite{2021MNRAS.508.3877G}

e\_MassH, Uncertainty on mass\_H (emass\_H), from \cite{2021MNRAS.508.3877G}

chisqH, chi square value of the fit (pure-H)(chisq\_H), from \cite{2021MNRAS.508.3877G}

TeffHe, Effective temperature from fitting the dereddened G, BP, and RP absolute fluxes with pure-He model atmospheres (teff\_He), from \cite{2021MNRAS.508.3877G}

e\_TeffHe, Uncertainty of teff\_He (eteff\_He), from \cite{2021MNRAS.508.3877G}

loggHe, Log of surface gravity from the dereddened G, BP, and RP absolute fluxes with pure-He model atmospheres (logg\_He) from \cite{2021MNRAS.508.3877G}

e\_loggHe, Uncertainty on logg\_He (elogg\_He), from \cite{2021MNRAS.508.3877G}

MassHe, Stellar mass resulting from the adopted mass-radius relation and best fit parameters (mass\_He), from \cite{2021MNRAS.508.3877G}

e\_MassHe, Uncertainty on mass\_He (emass\_He), from \cite{2021MNRAS.508.3877G}

chisqHe, chi square value of the fit (pure-He)(chisq\_He), from \cite{2021MNRAS.508.3877G}

fuvwd, Objects that appear to have a white dwarf in GALEX FUV-NUV

nuvwd, Objects that appear to have a white dwarf in GALEX NUV-Gaia Bp

gaiawd, Objects that appear to have a white dwarf in Gaia Bp-Rp

wd\_m\_H, Estimated white dwarf mass from fit of white dwarf with pure-H atmosphere.

wd\_m\_H\_e, Estimated white dwarf mass error for pure-H atmosphere

wd\_age\_H, Estimated white dwarf age from fit of white dwarf with pure-H atmosphere.

wd\_age\_H\_e, Estimated white dwarf age error for pure-H atmosphere

wd\_teff\_H, Estimated white dwarf temperature from fit of white dwarf with pure-H atmosphere.

sec\_m\_H, Estimated secondary non-degenerate from fit of white dwarf with pure-H atmosphere. If no evidence of a companion from the photometry, we performed single WD fit and report 0 for this value.

sec\_m\_H\_e, Estimated secondary non-degenerate mass error pure-H white dwarf atmosphere.

chi\_sq\_H, The resulting chi squared value for the fit for hydrogen atmospheres. 

goodness\_H, The probability that the estimated masses are a good fit for the data for hydrogen atmospheres.

wd\_m\_He, Estimated white dwarf mass from fit of white dwarf with pure-He atmosphere.

wd\_m\_He\_e, Estimated white dwarf mass error for pure-He atmosphere

wd\_age\_He, Estimated white dwarf age from fit of white dwarf with  pure-He atmosphere. 

wd\_age\_He\_e, Estimated white dwarf age error for pure-He atmosphere

wd\_teff\_He, Estimated white dwarf temperature from fit of white dwarf with  pure-He atmosphere. 

sec\_m\_He, Estimated secondary non-degenerate star from fit of white dwarf with pure-He atmosphere. If no evidence of a companion from the photometry, we performed single WD fit and report 0 for this value.

sec\_m\_He\_e, Estimated secondary non-degenerate mass error white dwarf pure-He atmosphere.

chi\_sq\_He, The resulting chi squared value for the fit for helium atmospheres. 

goodness\_He, The probability that the estimated masses are a good fit for the data for helium atmospheres.


\end{document}